\documentstyle[12pt,aasms4]{article}

\def\arcsecpoint{$''\!.$}
\received{}
\accepted{}

\slugcomment{submitted to {\it The Astrophysical Journal}}

\lefthead{Kraemer, Crenshaw, \& Gabel}
\righthead{STIS Echelle Observations of the Intrinsic UV Absorption in the 
Seyfert 1 Galaxy NGC 3783.\altaffilmark{1}}

\begin{document}

\title{STIS Echelle Observations of the Intrinsic UV Absorption in the 
Seyfert 1 Galaxy NGC 3783.}

\author{S. B. Kraemer\altaffilmark{2},
D. M. Crenshaw\altaffilmark{2},
\& J.R Gabel\altaffilmark{2}}

\altaffiltext{1}{Based on observations made with the NASA/ESA Hubble Space
Telescope. STScI is operated by the Association of Universities for Research in
Astronomy, Inc. under the NASA contract NAS5-26555. }
  
\altaffiltext{2}{Catholic University of America,
NASA's Goddard Space Flight Center, Code 681,
Greenbelt, MD  20771; stiskraemer@yancey.gsfc.nasa.gov, 
crenshaw@buckeye.gsfc.nasa.gov, gabel@iacs.gsfc.nasa.gov.}

\begin{abstract}

We present observations of the UV absorption lines in the Seyfert 1 galaxy 
NGC 3783, obtained with the 
medium resolution ($\lambda$/$\Delta\lambda$ $\approx$ 40,000) echelle gratings
of the Space Telescope Imaging Spectrograph (STIS) on the
{\it Hubble Space Telescope (HST)}. The spectra reveal the presence
of three kinematic components of absorption in Ly$\alpha$, C~IV, and N~V, at 
radial velocities of $-$1365, $-$548, and 
$-$724 km s$^{-1}$ with respect to the
systemic velocity of the host galaxy 
(Components 1, 2 and 3, respectively); Component 1
also shows absorption by Si~IV. Component 3 was not
detected in any of the earlier Goddard High Resolution Spectrograph (GHRS) 
spectra,
and the C~IV absorption in the other components has changed since the
most recent GHRS observation obtained $\sim$ 5 yr earlier. Somewhat
unexpectedly, each component has a covering factor (of the
continuum source $+$ broad emission line region) of $\sim$ 0.6. We have
calculated photoionization models to match the UV column densities in
each of the three components. The models predict a zone 
characterized by high ionization
parameter (U $=$ 0.65 -- 0.80) and column density (6.4 x 10$^{20}$
-- 1.5 x 10$^{21}$ cm$^{-2}$) for each component,
and a second, low ionization (U $=$ 0.0018) and low column density
(4.9 x 10$^{18}$ cm$^{-2}$) zone for Component 1. Based on the model
results, there should be strong absorption in the bandpass of the
{\it Far Ultraviolet Spectroscopic Explorer (FUSE)}, 912 \AA~ --
1200 \AA~,
including saturated O~VI lines at each component velocity. The models
also predict large O~VII and O~VIII column densities but 
suggest that the UV absorbers 
cannot account for all of the X-ray absorption detected in 
recent {\it Chandra} spectra. Finally, there is no evidence for a
correlation between the characteristics of the UV absorbers and the
UV continuum flux, and, by inference, the ionizing continuum. Hence, we 
suggest that the variations observed in the
GHRS and STIS spectra are due in a large part to changes in the column
densities of the absorbers as the result of transverse motion.

\end{abstract}


\section{Introduction}

Since the launch of the {\it International Ultraviolet Absorber (IUE)}, it has been
known that the UV spectra of Seyfert 1 galaxies show absorption 
lines intrinsic to their nuclei 
(Ulrich 1988). With the advent of the {\it HST},
it is now understood that intrinsic absorption is a common phenomenon,
present in more than half of the well-studied Seyfert 1 galaxies (Crenshaw
et al. 1999). Among those Seyferts that show absorption, high ionization lines
such as N~V $\lambda\lambda$1238.8, 1242.8
and C~IV $\lambda\lambda$1548.2, 1550.8 are always present, along with
Ly$\alpha$, while lower ionization lines, such as 
Si~IV $\lambda\lambda$1393.8, 1402.8 and Mg~II
$\lambda\lambda$2796.3, 2803.5, are less common. The absorption lines are 
blueshifted (by up to 2100 km s$^{-1}$) with respect to the
systemic velocities of the host galaxies, indicating net radial outflow. 
The ionic columns are highly variable, which may be the result
of changes in response to the ionizing continuum (cf. Krolik \&
Kriss 1997; Shields \& Hamann 1997) or transverse motion (Crenshaw \&
Kraemer 1999). In either case, the variability is indicative of
the promixity of the absorbers to the central active nucleus of these
galaxies.

The presence of intrinsic absorption, typically in the form of 
bound-free edges of O~VII and O~VIII, has been detected in the X-ray 
spectra of a similar fraction of Seyfert 1 galaxies (Reynolds 1997; 
George et al. 1998). Most recently, spectra obtained with the {\it Chandra
X-ray Observatory}
have revealed that X-ray absorption lines associated with this
material are also blue-shifted (Kaastra et al. 2000; Kaspi et al. 2000,
hereafter K2000). 
The connection between the X-ray and UV absorption is complex.
It has been argued that some fraction of the UV absorption
arises in the same gas responsible for the X-ray absorption (Mathur, Elvis
\& Wilkes 1995, 1999; Crenshaw \& Kraemer 1999; Kriss et al.
2000). Some sources, notably NGC 4151 (Kriss 1998; Kraemer et al. 2001)
and NGC 3516 (Kriss et al. 1996; Crenshaw, Maran, \& Mushotzky 1998) 
possess low ionization lines, indicative of a wide range
in physical conditions within the absorbers. Furthermore, there
is evidence that the X-ray absorbers in some sources must also
be multi-zoned (Otani et al. 1996; Reynolds et al. 1997).

NGC 3783 ($z$ $=$ 0.00976) is a bright, well-studied Seyfert 1 galaxy.
During an {\it IUE} monitoring campaign from 1991 Decemeber 21
to 1992 July 29, the UV continuum 
showed rapid (20 -- 40 days), large amplitude 
(factors of $\sim$ 2 at 1460 \AA~) flux variations (Reichert et al.
1994). Intrinsic Ly$\alpha$ and C~IV absorption were first detected in Faint
Object Spectrograph (FOS) spectra by Reichert et al. (1994), and
N~V absorption was first seen in GHRS spectra by Lu, Savage, \& Sembach
(1994). Subsequent GHRS spectra revealed the UV absorption lines to be
highly variable (Maran et al. 1996; Crenshaw 1999), with the C~IV line
becoming undetectable at one point (during observations taken in 1993
February 5). Observations with {\it ASCA} revealed the presence of
a large ($>$ 10$^{22}$ cm$^{-2}$) and variable column of ionized gas (George et al.
1998). Using photoionization models, with parameters derived from the 
fit to the X-ray absorber by George,
Turner, \& Netzer (1995), Shields \& Hamann (1997)  
suggested that was plausible that the X-ray and UV absorption arose in
a single zone, and that the changes in the absorber were due to
the variable luminosity of the ionizing continuum. From spectra obtained with the 
High Energy Transmission Grating Spectrometer aboard {\it Chandra},
K2000 derived a mean velocity for the X-ray lines of
$\approx$ 440 $\pm$ 200 km s$^{-1}$, which is consistent with the strong UV 
absorber discussed by Maran et al. (1996). Although this seems to support
the suggestion that the absorbers are associated, the ionization state
of the X-ray absorber observed by K2000 appears to be too
high to produce the observed UV lines (although this depends on the
spectral energy distribution [SED] of the ionizing radiation emitted by
the central source). Interestingly, the total column of O~VII K2000
derived from photoionization modeling of the absorption line gas is roughly an order of 
magnitude less than estimated using {\it ASCA} data
(Reynold et al. 1997; George et al. 1998). Hence, the exact nature of the
absorption in NGC 3783 is, as yet, undetermined.

NGC 3783 is one of the targets, along with NGC 5548 (Crenshaw \& Kraemer 1999),
NGC 4151 (Crenshaw et al. 2000; Kraemer et al. 2001) and NGC 3516
(Crenshaw \& Kraemer, in preparation), in STIS Instrument Definition
Team programs to investigate intrinsic absorption in Seyfert galaxies
using medium resolution echelle
spectra. In this paper, we present our analysis of the intrinsic
absorber in NGC 3783, based on the echelle spectra and photoionization
models.
The paper is organized as follows: in Section 2 we describe the observations, 
in Section 3, we give the details of the measurement of the intrinsic lines, 
in Sections 4 and 5 we detail the photoionization modeling of the 
absorbers, and in Section 6 we discuss the implications of the
results.

\section{Observations}

We obtained STIS medium-resolution echelle spectra of the nucleus of NGC 3783
through a 0\arcsecpoint2 x 0\arcsecpoint2 aperture on 2000 February 27. These 
observations, along with previous observations of the N~V and C~IV regions with 
the GHRS, are described in Table 1. The 
reduction of the GHRS spectra are described in Maran et al. (1996) and Crenshaw 
et al. (1999). We reduced the STIS echelle spectra using the IDL software 
developed at NASA's Goddard Space Flight Center for the STIS Instrument 
Definition Team.
The data reduction includes a procedure to remove the background light from each 
order using a scattered light model devised by Lindler (1999). The individual 
orders in each echelle spectrum were spliced together in the regions of overlap.

Figure 1 shows a comparison of the GHRS and STIS spectra in the C~IV region. As 
noted in Crenshaw et al. (1999), NGC~3783 shows dramatic variability in the 
column densities of the intrinsic C~IV absorption lines. The GHRS spectrum on 
1993 February 5 shows only Galactic C~IV and C~I lines (see Lu et al. 1994, 
Maran et al. 1996), and no evidence for intrinsic C~IV absorption. However, 11 
months later a C~IV $\lambda\lambda$1548.2, 1550.8 doublet (component 2) appears 
at a radial velocity of $-$548 km 
s$^{-1}$ (relative to the systemic redshift of z $=$ 0.00976). After a 
subsequent interval of 15 months, another doublet (component 
1) appears at $-$1365 km s$^{-1}$, and the original doublet is still present at 
about the same strength. Interestingly, just 16 days after no C~IV absorption 
was detected in 1993, a GHRS spectrum shows the N~V $\lambda\lambda$1238.8, 
1242.8 doublet in absorption at the radial velocity of component 2.

Our STIS observations, obtained almost five years after the last GHRS spectrum, 
show more changes in the C~IV absorption. The column density of 
component 1 has increased tremendously and a third C~IV doublet has 
appeared at $-$724 km s$^{-1}$. This component (3) is very strong and blended 
with the others, so that careful deblending is required to determine the column 
densities of the individual components.
\footnote{The numbering of components is not intuitive. Crenshaw et al. (1999) 
numbered the first two components in order of increasing wavelength; however 
component 3 has now appeared in between these two.}

Figure 2 shows the regions in the STIS E140M spectra where intrinsic absorption 
lines were detected (L$\alpha$, N~V, C~IV, and Si~IV). Fluxes are 
plotted as a function of the radial velocity of the strongest member of each 
doublet, relative to the systemic redshift.
Each of the three absorption components are present in L$\alpha$ and the 
doublets of C~IV and N~V.
Although it is difficult to distinguish component 2 in N~V and 
C~IV due to blending with component 3, its presence in 
L$\alpha$ shows that this component has not disappeared at this epoch. 
Only component 1 is present in Si~IV $\lambda\lambda$1393.8, 1402.8, indicating 
the presence of lower ionization gas in this component.
Examination of the E230M spectrum reveals no discernable intrinsic absorption 
for any ion in this spectrum, including Mg~II.

\section{Measurements}

The procedures we used to measure the intrinsic absorption lines 
follow those of Crenshaw et al. (1999). To determine the shape of the underlying 
emission, we fit a cubic spline to regions on either side of the absorption. To 
normalize the absorption profiles, we divided the 
observed spectra by the spline fits. 
We consider the case where an absorber does not completely cover the emission 
sources behind it. If this effect is present and not corrected for, the column 
densities will be underestimated. 
We must therefore determine the covering factor C$_{los}$, which is the fraction 
of continuum plus BLR emission that is occulted by the absorber in our line of 
sight. We consider the case where the covering factor differs from one kinematic 
component to the next, but not within a component, as the signal-to-noise is not 
sufficient to determine the covering factor across the profiles. 
The value of C$_{los}$ for a component can be determined from a doublet 
(Hamann et al. 1997). If the expected ratio of the 
optical depths of the doublet lines is 2 (as for C~IV, N~V, and Si~IV), then
\begin{equation}
C_{los} = \frac{I_1^2 - 2I_1 + 1}{I_2 - 2I_1 + 1},
\end{equation}

where I$_1$ and I$_2$ are the residual fluxes in the cores of the weaker line 
(e.g., N~V $\lambda$1242.8) and stronger line (e.g., N~V $\lambda$1238.8), 
respectively.

As can be seen in Figure 2, the N~V components are easier to isolate than those 
in C~IV, and we therefore used the N~V doublet to determine C$_{los}$.
Table 2 gives C$_{los}$ for each kinematic component, along with the radial 
velocity centroid and FWHM derived from the isolated components (see below). 
The covering factors for all three components are close to 0.6. By contrast, 
four other Seyfert 1 galaxies observed at high dispersion by {\it HST} have at 
least one component with C$_{los}$ $>$ 0.9 (Crenshaw et al. 1999).

Since the absorption components are resolved, we can determine the column 
density of each component from its optical depth as a function of radial 
velocity ($v_r$). The optical depth at a particular radial velocity is:
\begin{equation}
\tau = ln \left(\frac{C_{los}}{I_r + C_{los} -1}\right) 
\end{equation}
(Hamann et al. 1997).

The ionic column density is then obtained by integrating the optical depth 
across the profile:
\begin{equation}
N = \frac{m_{e}c}{\pi{e^2}f\lambda}~\int \tau(v_{r}) dv_{r},
\end{equation}
(Savage \& Sembach 1991), where $f$ is the oscillator strength and $\lambda$ is 
the laboratory wavelength (Morton et al. 1988).

The kinematic components are blended in many cases, and we therefore used the 
following procedure to deblend the components and determine their ionic column 
densities. 
Since the absorption components are additive in optical depth, we converted the 
normalized absorption for each ion to optical depth (equation 2) as a function 
of radial velocity.
For C~IV, we derived the optical depth profile for component 1 directly from the 
$\lambda$1548.2 line in the STIS data (Figure 1). For C~IV component 2, we used 
the combined $\lambda$1550.8 profile from the 1994 and 1995 GHRS observations, 
which is also unaffected by contaminating intrinsic or Galactic absorption. To 
obtain the profile for C~IV component 3, we subtracted the component 1 doublet 
from the optical depth blend in the STIS data (given the 2:1 for the doublet).
The profile for component 2 was obtained at a different time, so we 
scaled and subtracted the component 2 doublet until a reasonably 
smooth profile was obtained for component 3. 
With the optical depth profiles for each component, we then repeated the 
deblending process for the GHRS data until only the Galactic C~I lines remained
in this region.

For the N~V components, we used a similar scheme. We obtained the optical depth 
profile of component 1 directly from the N~V $\lambda$1238.8 line in the STIS 
data, and component 2 from the same line in the 1993 GHRS observation (see 
Crenshaw et al. 1999). The profiles of components 1 and 2 were then reproduced 
at the positions of the doublets in the STIS data, and scaled and subtracted 
until a smooth profile was obtained for component 3. The Si~IV lines for 
component 1 are unaffected by contaminants, so we determined the optical depth 
profile directly from the STIS data. 
Errors in the optical depths were determined from the photon noise (which 
dominated in Si~IV) and the uncertainties in scale factors used in the 
deblending process (which dominated in C~IV and N~V). For each kinematic 
component, the optical depth profiles for different ions as a function of radial 
velocity agree well; there is no evidence for a changing ionic ratio across any 
component. The small residuals obtained from the deblending process validate our 
assumption that the covering factors derived from N~V are approximately correct 
for C~IV and Si~IV.

Table 2 gives the radial velocity centroids, widths (FWHM), and associated 
one-sigma errors for each kinematic component. These values represent the means 
and standard deviations determined from the measurements of the C~IV and N~V 
optical depth profiles (and the Si~IV profile
for component 3). As in previous studies (Crenshaw et al. 1999), we find no 
evidence for changes in velocity centroids or widths over time.
In Table 3, we give the ionic column densities for each component, which were 
derived from the integration of optical depths according to equation 3.
The Ly$\alpha$ absorption in each component appears to be heavily saturated, 
given the evidence for covering factors substantially smaller than one; we quote 
only lower limits for the H~I in these components, derived from the C$_{los}$ 
$=$ 1 case. We have also determined upper limits to the columns of C~II, Si~II, 
and Mg II, which are given in Table 3.

\section{Inputs to the Photoionization Models}

 Photoionization models for this study were generated using the
 code CLOUDY90 (Ferland et al. 1998). We have modeled the absorbers
 as matter-bounded slabs of atomic gas, irradiated
 by the ionizing continuum radiation emitted by the central source.
 As per convention, the models
 are parameterized in terms of the ionization parameter,
 U, the ratio of the density of photons with energies $\geq$ 13.6 eV
 to the number density of hydrogen atoms at the illuminated face of 
 the slab. Each separate kinematic component was initially modeled with one
 set of initial conditions, i.e., U, n$_{H}$, and the total
 hydrogen column density, N$_{H}$ ($=$ N$_{H~I}$ $+$ N$_{H~II})$\footnote{We 
 use N(XM) to denote the ionic column density, where ``X'' is the atomic 
 symbol and ``M'' is the ionization state.}. 
 As discussed in Section 5, a second model (or ``zone'') was required to fit
 Component 1. For the models, we have assumed only thermal 
 broadening, since 1) the absorption line widths could be due to the
 superposition of unresolved kinematic components and 2) comparison models 
 assuming turbulent velocities
 of $\leq$ 300 km s$^{-1}$ predicted nearly identical ionic columns.
 As in similar studies (e.g. Crenshaw
 \& Kraemer 1999), a model is deemed successful when the
 predicted ionic column densities, specifically those of N~V, C~IV and 
 (in one case) Si~IV, provide a good match (i.e., better than
 a factor of 2) to those observed in the STIS spectra.

 In order to approximate the SED of the
 ionizing continuum, we assumed it to be a power-law of the form
 F$_{\nu}$ $\propto$ $\nu^{-\alpha}$. Based on {\it ASCA} observations, 
 George et al. (1998) fit the 0.6 -- 10 keV with a power-law of index
 $\alpha$ $\approx$ 0.8, with an average flux in the 0.1 - 10 keV band of 
 $\sim$
 2.8 x 10$^{-10}$ ergs s$^{-1}$ cm$^{-2}$ (which yields F$_{\nu}$ $\sim$ 
 3.5 x 10$^{-28}$ ergs s$^{-1}$ cm$^{-2}$ Hz$^{-1}$ at 0.6 keV).
 In our STIS spectra, the flux at 1470 \AA~ is F$_{\nu}$ $\approx$ 
 2.53 x 10$^{-26}$ ergs s$^{-1}$ cm$^{-2}$ Hz$^{-1}$. In order to determine
 the flux at the Lyman limit, we extrapolated
 to 912 \AA~ assuming a power-law with a spectral index of $\alpha$ $=$ 1,
 which yields a flux of F$_{\nu}$ $\sim$ 7.5 x 10$^{-26}$ ergs s$^{-1}$ cm$^{-2}$ 
 Hz$^{-1}$, after correcting for reddening (E$_{B-V}$ $=$ 0.12; 
 Reichert et al. 1994). Extending the X-ray continuum to UV wavelengths
 underpredicts the observed UV flux by more than two orders of
 magnitude and, therefore, the continuum must steepen
 at energies below 0.6 keV. For the sake of simplicity, we assumed that the 
 break in the
 continuum occurs at this energy. Therefore, we assume the following
 UV to X-ray spectral indices: $\alpha = 1$ below 13.6~eV, 
 $\alpha = 1.4$ over the range 13.6~eV $\leq h\nu <$ 600~eV, 
 and $\alpha = 0.8$ above 600~eV. It should be noted that the
 power-law break could occur at lower energies and/or that a ``Big Blue
 Bump'' may be present (see George et al. 1995). If the former were the case, 
 the relative fractions of highly ionized species, such as O~VII and O~VIII, 
 would be somewhat higher
 than our predictions, while in the latter case the opposite would be
 true. The lack of simultaneous UV and X-ray observations 
 introduces an additional uncertainty in modeling the SED.
  
 We have assumed roughly solar element abundances (cf. Grevesse \& Anders 1989),
 which are, by number relative to H, as follows: He $=$ 0.1, 
 C $=$ 3.4 x 10$^{-4}$, N $=$ 1.2 x 10$^{-4}$, O $=$ 6.8 x 10$^{-4}$,
 Ne $=$ 1.1 x 10$^{-4}$, Mg $=$ 3.3 x 10$^{-5}$, Si $=$ 3.1 x 10$^{-5}$,
 S $=$ 1.5 x 10$^{-5}$, and Fe $=$ 4.0 x 10$^{-5}$. The absorbing gas
 is assumed to be free of cosmic dust.

\section{Model Results}

Component 1, the most blueshifted of the absorbers, is the only one 
to possess detectable Si~IV absorption. The ratio of N(N~V)/N(C~IV)
is $>$ 2, which is indicative of highly ionized gas (see, for example,
Shields \& Hamann 1997), while, on the other hand, the fact that N(C~IV)/N(Si~IV) 
is smaller than the abundance ratio is evidence for much lower ionization
gas. Based on our experience modeling the absorbers in NGC 4151 (Kraemer et al.
2001), this dilemma can be resolved if the ionizing continuum were
modified by an intervening absorber which is optically thick
from 13.6 eV -- 100 eV, however there is no evidence for such material
in our line-of-sight to the nucleus of NGC 3783. 
A simpler explanation is that Component 1 consists of two zones characterized
by different ionization parameters, with similar covering factors and
velocity profiles; in this scenario, the lower ionization absorption 
could arise in gas co-located with the more tenuous, and thus more highly ionized,
absorber. We modeled Component 1 using two separate zones with the following
parameters: 1) U$=$ 0.78, N$_{H}$ $=$ 1.2 x 10$^{21}$ cm$^{-2}$, and
2) U $=$ 0.0018, N$_{H}$ $=$ 4.9 x 10$^{18}$ cm$^{-2}$. The 
predicted ionic columns are listed in Tables 4 and 5, respectively (the
ions used to constrain the models are shown in boldface). If the
two zones are at the same radial distance, the density of the lower
ionization zone is $\sim$ 430 times higher. 
The sums of the predicted ionic 
columns from the two zones provide a good fit to the observed
columns. Also, the low ionization model predictions for N(C~II) and N(Si~II),
are below their respective 2$\sigma$ limits, which agrees with the
non-detection of such lines in the STIS spectra. It should be noted that the
models slightly underpredict N(Si~IV) and overpredict
N(C~IV). \footnote{The only way to increase N(Si~IV), given the absence of 
lower ionization
lines, is to increase both the ionization parameter and column density of 
the low ionization zone model. However, even slight changes ($\sim$ 20\%)
in these parameters would worsen the N(C~IV) prediction. We will be
better able constrain the range in physical conditions in Component 1 with
the series of observations taken during 
our monitoring campaign (see Section 7).} 

Based on the ratio of the N~V and C~IV column densities ($\sim$ 5.8), Component 2 
is the most highly ionized of the 3 kinematic components detected in the
STIS spectra. As shown in Table 6, we were able to fit the observed
ionic columns with a single-zoned model, parameterized as follows:
U $=$ 0.80 and N$_{H}$ $=$ 6.4 x 10$^{20}$ cm$^{-2}$. Nevertheless,
due to the large uncertainties in the measured columns (see Table 3),
the physical conditions in Component 2 are not tightly constrained.
For example, if the N~V to C~IV ratio were $\sim$ 10 (which is well within the
errors), the ionization parameter would be $\sim$ 2.0 and 
N$_{H}$ $\sim$ 10$^{22}$ cm$^{-2}$, which would yield an order of 
magnitude increase in O~VII and O~VIII columns. 

While Component 3 possesses the strongest N~V absorption, the N(N~V)/N(C~IV) 
ratio is less than that of Component 2, indicative of a lower
ionization parameter. As shown in Table 7, we were able to match the observed column densities
with a single-zoned model parameterized as follows: U $=$ 0.65 and
N$_{H}$ $=$ 1.5 x 10$^{21}$ cm$^{-2}$.

In summary, we have been able to fit the measured ionic column densities
with simple one- or two-zoned photoionization models. There is no evidence
for anomalous abundances or cosmic dust (i.e., depletions of elements from gas phase) 
within the absorbers. None of the absorbers possesses sufficient
opacity at the H~I or He~II Lyman limits to affect the ionization state
of gas further from the nucleus (for a discussion of this effect, see
Kraemer et al. 1999), hence we see no effects due to the screening
of one absorbing component by another.

\section{Discussion}

\subsection{X-ray and far-UV Absorption}

Based on our model predictions, all three components should possess substantial
columns of O~VII, and O~VIII. The summed O~VII and O~VIII columns
are 1.3 x 10$^{18}$ cm$^{-2}$ and 7.7 x 10$^{17}$ cm$^{-2}$, respectively.
The predicted O~VIII column is 5 times less than that determined by 
K2000, based on modeling of roughly contemporary {\it Chandra} 
spectra, while 
the predicted O~VII column is a factor of 2.6 greater. 
Although the predicted and observed
column densities do not agree, the radial velocities of Components 2 and 3 
are roughly consistent with the mean value of the X-ray absorption lines 
(K2000). If the Component 2 is more highly ionized than indicated 
by the mean N~V to C~IV ratio, its O~VIII column may be significantly
larger. However,
this would worsen the fit for O~VII and the model would 
still underpredict the columns of the high 
ionization species detected
by K2000, such as Si~XIV, Fe~XVIII, and Fe~XIX. Since the
N~V and C~IV columns in Component 3 have much smaller errors, the ionization
parameter is well constrained by the models, hence the absorber cannot
produce the observed O~VIII column. Interestingly, our model for
Component 1 predicts an O~VII $\lambda$18.6 line with an $EW$ $=$ 26 m\AA~ and
O~VIII $\lambda$16.0 line with an $EW$ $=$ 28 m\AA~. Although not
reported by K2000, perhaps
these feature are present on the blue wings of the lines detected in the
{\it Chandra} spectra.

To summarize, although slight variations in
our model parameters could bring the O~VII prediction into
agreement with the {\it Chandra} results, a substantial
fraction of the X-ray absorbing gas is too highly ionized to 
produce UV absorption. Also, the O~VII column reported by
K2000 is significantly
less than that determined from previous {\it ASCA} results (Reynolds 1997;
George et al. 1998) which implies that the either 1) the absorbing column 
was larger in the past, or 2) there are unknown problems in 
determining the O~VII column from one of these datasets. 
In either case, we cannot, with sufficient certainty, establish 
a one-to-one connection between the X-ray and UV absorbers.

In Table 8 we list the line center
optical depths, based on our model predictions, for several far-UV lines
which will be detectable with {\it FUSE}. 
Clearly, the O~VI $\lambda\lambda$1031.9, 1037.6 lines will be saturated 
($\tau_{center}$ $>$ 4) for
all three kinematic components, and Component 1 should possess
saturated C~III $\lambda$997 and strong N~III $\lambda$990.
Although, as noted in Section 2, Ly$\alpha$ is saturated in each component, 
our results indicate that Ly$\beta$ and Ly$\gamma$ may not be in
Components 2 and 3. Hence 
these lines 
could be used to determine the total H~I column, which, in turn, can be 
used to test our model assumptions (and better constrain the physical 
conditions in Component 2).

\subsection{Changes in the UV Absorbers}

Although the column densities of the UV absorbers in NGC 3783 (and other 
Seyfert 1s) are known to change over time (see Crenshaw et al. 1999), these
variations could be due to the response of the absorbers to changes in the
ionizing flux, changes in total column due to transverse motion, or 
some combination of these effects. In NGC 4151, Kraemer et al. (2001)
argued that the largest changes in the UV absorbers were well-correlated
with changes in the ionizing continuum, while for NGC 5548, Crenhsaw \& 
Kraemer (1999) found that transverse motion was a more plausible explanation. 
Shields \& Hamann (1997) suggested that the variations in UV absorption 
observed in NGC 3783 are 
a function of changes in the luminosity of the ionizing source. Here, we
re-examine this suggestion, using an additional GHRS observation (from 1995 April
11) along with our 
STIS dataset (we have not included the FOS data from 1992 July 27 in
this analysis, due to the unreliability of ionic column density measured in
low resolution spectra).

Figure 3 shows the changes in the
the C~IV columns,  
assuming that undetectable columns are
$<$ 3 x 10$^{13}$ cm$^{-2}$, and the UV continuum
over time. First, the C~IV columns do not seem to be correlated
with the UV flux. For Component 2, the C~IV is 
undetectable in 1993 and strong in 1994, during which time the continuum flux 
decreases, and the column decreases from 1995 to 2000 while the flux increases.
Although these column changes could be interpreted as a response
to the continuum flux, this cannot also explain the
the weak response between 1994 and 1995, during which time the UV flux
has dropped sharply. Furthermore, Component 3 would have to be undetectable
at both low and high ionization states to have missed detection in earlier
epochs. Given our two-zone model for Component 1, it is difficult to produce 
the increase in
the C~IV column from 1995 to the present solely by increasing the ionizing
flux, since most of the gas is in the high ionization zone. We have also determined the ionization parameters that describe the 
physical state of the absorbers at each epoch, assuming that N$_{H}$ is 
constant (i.e., no transverse motion). As shown in Figure 3, the variations 
are not correlated with the 
changes in the UV flux. For example, there is no evidence that U is
at a mininum for any of the three components during 1995, when the UV flux
was weakest. Based on this, we see no evidence that the state
of the UV absorbers is simply the result of changes in the ionizing flux.

Alternatively, much of the observed variations in UV ionic columns could
be explained by transverse motion. For example, assuming constant U, the different C~IV 
columns observed in Component 2 correspond to values of N$_{H}$ ranging from 
3.2 x 10$^{20}$ cm$^{-2}$ (in 1993) to 1.5 x 10$^{21}$ cm$^{-2}$ (in 1994).
The predicted C~IV column in 1993 would be 3.0 x 10$^{13}$ cm$^{-2}$, which is 
at the detectability limit and could explain the absence of a 
noticeable feature.
For Component 1, the smaller C~IV column observed 
in 1995 could be simply due to the absence of the low ionization zone. 
Similarly, the absence of Component 1 in earlier epochs could be the result
of a smaller column of the high ionization gas, as may be the case
for Component 3. 

Although the changes in the ionic columns could be due solely to transverse
motion, we cannot rule out a combination of a changes in both total column 
density and ionization state. Deconvolving these effects requires
an intensive monitoring program, rather than the poor sampling
provided by the few available observations. Unfortunately, if the
variations are primarily due to transverse motion, one cannot rely
on recombination/ionization timescales
to constrain the densities of the absorbers.

\section{Summary}

We have used medium resolution echelle spectra obtained with {\it HST}/STIS
to study the physical conditions in the intrinsic UV absorbers in the
Seyfert 1, NGC 3783 and have determined the following.

1. The appearance of the UV absorption has changed noticeably since the
most recent GHRS observations (taken in 1995). There are currently
three distinct kinematic components, at 
velocities of $-$1365, $-$548, and $-$724 km s$^{-1}$, with respect to the systemic velocity of the host galaxy.
Each component possesses C~IV $\lambda\lambda$1548.2, 1550.8,
N~V $\lambda\lambda$1238.8, 1242.8, and heavily saturated Ly$\alpha$ lines,
while Component 1 also shows the presence of Si~IV $\lambda\lambda$1393.8,
1402.8. Component 3, which has the largest N~V column density, was 
absent in all earlier 
observations. As shown in Table 2, each component has a covering
factor of less than unity. The simplest explanation is that the
absorbers do not cover the entire broad line region, but other
geometries cannot be ruled out. 
In order to constrain the ionization
state and total hydrogen column densities of the absorbers, we used
photoionization models predictions to match the
measured ionic column densities. Components 2 and 3 were well
matched by single zoned models, while it required two zones of
different ionization parameter to fit Component 1. The ionization
parameters and hydrogen column densities were somewhat higher than
those observed in NGC 5548, with the exception of the one component
in that galaxy which may be associated with the X-ray absorption (Crenshaw
\& Kraemer 1999).

2. In contrast with Shields \& Hamann (1997), we do not find strong evidence
that the changes in the UV absorption are a function of variations
in the ionizing continuum. In fact, we find that these variations
may be more readily explained by transverse motion, which results in
changes the total
column of gas along our line-of-sight. Nevertheless, it is 
possible that both continuum variations and transverse motion
contribute to the variability of the absorption.

3. We predict that far-UV, {\it FUSE} spectra will show O~VI 
$\lambda\lambda$1031.9, 1037.6 absorption lines 
at the velocities
of the each UV kinematic component, and C~III $\lambda$977 and N~III $\lambda$990 at the
velocity of Component 1. The higher order Lyman lines
will not be saturated for Components 2 and 3, and, hence can be
used to check our model predictions. Although the photoionization models 
predict substantial O~VII and
O~VIII columns, and there is an apparent coincidence
between the velocity of UV Component 2 and the mean velocity of the
X-ray absorption lines, the absorbing gas is at a lower ionization state and 
of smaller total column than the gas responsible for the much of the X-ray 
absorption seen in {\it ASCA} (Reynolds 1997; George et al 1998) and {\it Chandra}
(K2000) spectra. Hence, it is likely that the X-ray
absorber consists of two of more zones, characterized by different physical
conditions. The determination
of the detailed relationship between the UV and X-ray absorption, and the
nature of their variability, will require simultaneous observations
and better sampling, which are the principal goals of an 
{\it HST}, {\it FUSE}, and {\it Chandra} monitoring campaign.

\acknowledgments

 S.B.K. and D.M.C. acknowledge support from NASA grant NAG5-4103. 
 We thank Ian George for illuminating discussions, and
 Wayne Landsman for help in the calibration of the STIS echelle spectra.

\clearpage

\figcaption[fig1.eps]{GHRS and STIS spectra of the C~IV region in NGC~3783. The 
Galactic lines are labeled in the top spectrum, and are due primarily to C~IV 
$\lambda\lambda$1548.2, 1550.8 and C~I $\lambda$1560.3 near the local standard 
of rest and C~I $\lambda$1560.3 from a Galactic high-velocity cloud at $+$240 km 
s$^{-1}$ (Lu et al. 1993; Maran et al. 1996). The kinematic components of the 
intrinsic C~IV absorption doublets are numbered. Dotted lines show the
zero flux levels.}

\figcaption[fig2.eps]{Portions of the STIS echelle spectra of NGC~3783, showing 
the intrinsic absorption lines detected. Fluxes are plotted as a function of the 
radial velocity (of the strongest member, for the doublets), relative to an 
emission-line redshift of z $=$ 0.00976. The kinematic components are identified 
with large and small numbers, for the strong and weak members of the doublets 
respectively. Absorption lines that are not labeled are Galactic.
The ``broad absorption'' feature seen in the Si~IV 
profile near $+$2800 km s$^{-1}$ is an artifact due to the STIS FUV MAMA 
repeller wire.}

\figcaption[fig3.eps]{Continuum and C~IV absorption variations in {\it HST}
high-resolution spectra as a function of Julian date. The top plot shows the
continuum variations at 1470 \AA\ in units of 10$^{-14}$ ergs s$^{-1}$ cm$^{-2}$
\AA$^{-1}$. The middle plot show variations in the C IV column density 
(Table 3)
in units of 10$^{14}$ cm$^{-2}$ for each component (1, 2, and 3). The lower plots shows ionization parameters 
(U)
derived from photoionization models assuming constant hydrogen column density,
as described in the text.}

\begin{deluxetable}{ccccll}
\tablecolumns{6}
\footnotesize
\tablecaption{{\it HST} High-Resolution Spectra of NGC 3783\label{tbl-1}}
\tablewidth{0pt}
\tablehead{
\colhead{Instrument} & \colhead{Grating} & \colhead{Coverage} &
\colhead{Resolution} & \colhead{Exposure} & \colhead{Date} \\
\colhead{} & \colhead{} & \colhead{(\AA)} &
\colhead{($\lambda$/$\Delta\lambda$)} & \colhead{(sec)} & \colhead{(UT)}
}
\startdata
STIS &E140M &1150 -- 1730 &46,000 &5400   &2000 February 27 \\
STIS &E230M &2274 -- 3119 &30,000 &2191   &2000 February 27 \\
& & & & \\
GHRS &G160M &1527 -- 1597$^{a}$ &20,000 &4134$^{b}$    &1993 February 5 \\
GHRS &G160M &1232 -- 1269$^{c}$ &20,000 &8064          &1993 February 21 \\
GHRS &G160M &1527 -- 1597$^{a}$ &20,000 &4788$^{b}$    &1994 January 16 \\
GHRS &G160M &1527 -- 1597$^{a}$ &20,000 &4352$^{b}$    &1995 April 11
\tablenotetext{a}{Includes the C~IV region (Maran et al. 1996; Crenshaw et al. 
1999).}
\tablenotetext{b}{per grating setting; two settings were used.}
\tablenotetext{c}{Includes the N~V region  (Lu et al. 1994; Crenshaw et al. 
1999.)}
\enddata
\end{deluxetable}

\clearpage
\begin{deluxetable}{cccr}
\tablecolumns{4}
\footnotesize
\tablecaption{Absorption Components in NGC~3783}
\tablewidth{0pt}
\tablehead{
\colhead{Component} & \colhead{Velocity$^{a}$} & \colhead{FWHM} & 
\colhead{C$_{los}$$^{b}$}\\
\colhead{} &\colhead{(km s$^{-1}$)} &\colhead{(km s$^{-1}$)} & \colhead{}
}
\startdata
1     &$-$1365 ($\pm$20) & 193 ($\pm$17) & 0.61 ($\pm$0.04) \\
2     &$-$548 ($\pm$15)  & 170 ($\pm$13) & 0.60 ($\pm$0.05) \\
3     &$-$724 ($\pm$15)  & 280 ($\pm$30) & 0.65 ($\pm$0.03) \\
\tablenotetext{a}{Velocity centroid for a systemic redshift of z $=$ 0.00976.}
\tablenotetext{b}{Covering factor in the line of sight from the N V doublet.}
\enddata
\end{deluxetable}

\begin{deluxetable}{lcccc}
\tablecolumns{5}
\footnotesize
\tablecaption{Ionic Column Densities (10$^{14}$ cm$^{-2}$)$^{a}$}
\tablewidth{0pt}
\tablehead{
\colhead{Date (UT)} & \colhead{Ion} & \colhead{Component 1} &
\colhead{Component 2} & \colhead{Component 3} }
\startdata
1993 Feb. 5, 21   &C~IV  &-----            &-----             &----- \\
              &N~V   &-----            &1.84 ($\pm$0.19)  &----- \\
1994 Jan. 16  &C~IV  &-----            &1.48 ($\pm$0.29)  &----- \\
1995 Apr. 11  &C~IV  &2.09 ($\pm$0.34) &1.24 ($\pm$0.23)  &----- \\
2000 Feb. 27  &C~IV  &3.50 ($\pm$0.34) &0.63 ($\pm$0.24)  &3.23 ($\pm$0.48) \\
              &N~V   &7.60 ($\pm$1.50) &3.68 ($\pm$1.84)  &16.03 ($\pm$2.32) \\
              &Si~IV &0.61 ($\pm$0.19) &-----             &----- \\
              &H~I   &$>$ 0.84 &$>$0.77 &$>$3.86 \\
              &C~II  &$<$1.0 &$<$1.0 &$<$1.0 \\
              &Si~II &$<$0.3 &$<$0.3 &$<$0.3 \\
              &Mg~II &$<$0.1 &$<$0.1 &$<$0.1
\tablenotetext{a}{quoted errors are 1$\sigma$ values.}
\enddata
\end{deluxetable}

\clearpage
\begin{deluxetable}{llll}
\tablecolumns{4}
\footnotesize
\tablecaption{Predicted Line Center Optical Depths for far-UV lines}
\tablenum{8}
\tablewidth{0pt}
\tablehead{
\colhead{Ion} &
\colhead{Line} &
\colhead{N (cm$^{-2}$)} & 
\colhead{$\tau_{center}$}
}
\startdata
 & Component 1 & (FWHM $=$ 193 km s$^{-1}$) & \\
\hline
H~I & Ly$\alpha$ & 2.0 x 10$^{16}$ & 130 \\
    & Ly$\beta$ &                  & 21 \\
    & Ly$\gamma$ &                 & 7 \\
C~III & $\lambda$977 & 1.2 x 10$^{15}$ & 11 \\
N~III & $\lambda$990 & 4.2 x 10$^{14}$ & 0.57 \\
O~VII & $\lambda$1032 & 2.6 x 10$^{16}$& 44 \\
\hline
 & Component 2 & (FWHM $=$ 170 km s$^{-1}$) & \\
\hline
H~I & Ly$\alpha$ & 1.7 x 10$^{15}$ & 13 \\
    & Ly$\beta$ &                  & 2.0 \\
    & Ly$\gamma$ &                 & 0.7 \\
O~VII & $\lambda$1032 & 1.3 x 10$^{16}$ & 25 \\
\hline
 & Component 3 & (FWHM $=$ 280 km s$^{-1}$) & \\
\hline
H~I & Ly$\alpha$ & 5.5 x 10$^{15}$ & 25 \\
    & Ly$\beta$ &                  & 4 \\
    & Ly$\gamma$ &                 & 1.4 \\
O~VII & $\lambda$1032 & 4.9.3 x 10$^{16}$ & 58 \\
\enddata
\end{deluxetable}

\clearpage
\begin{figure}
\plotone{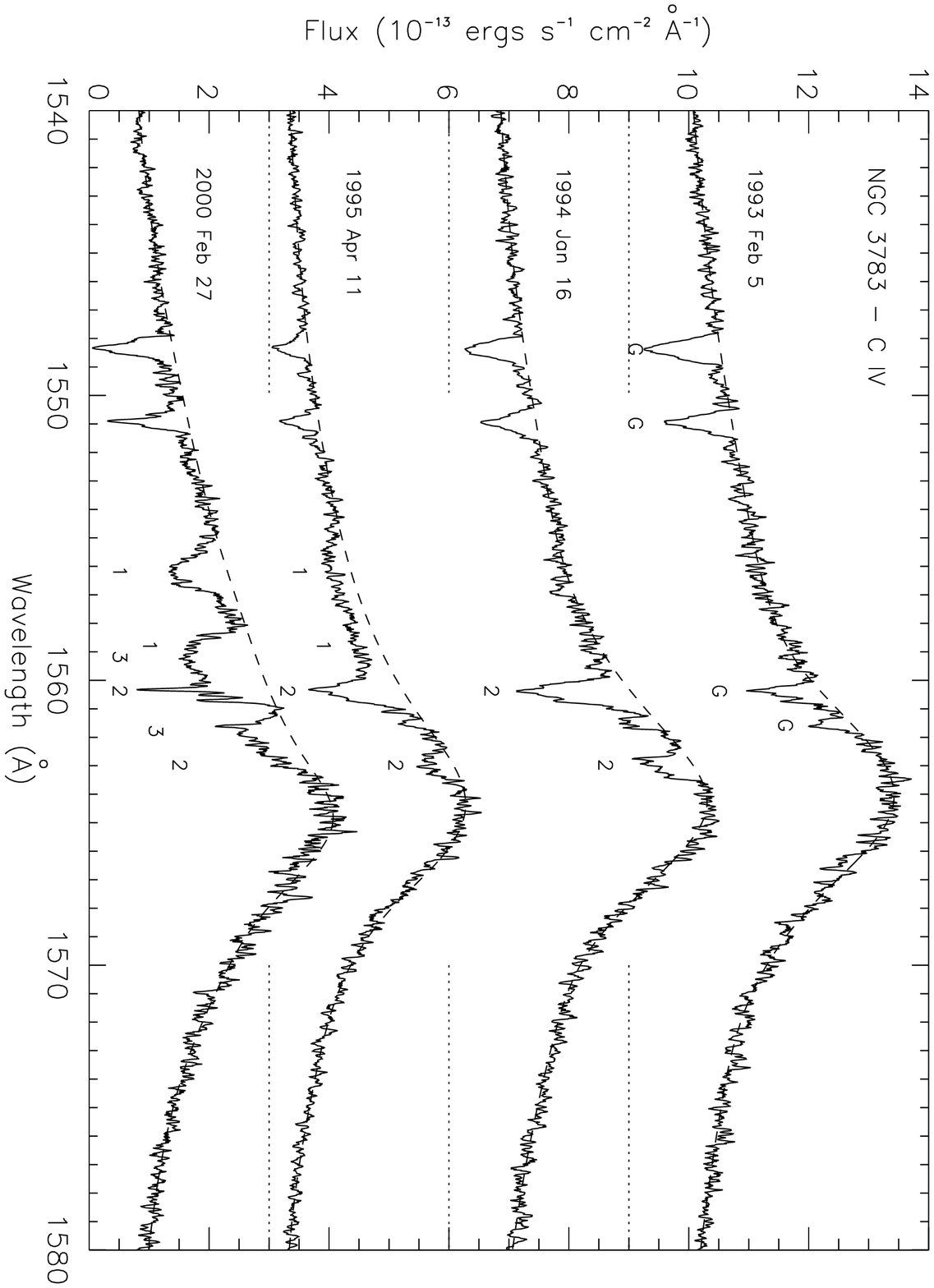}
\\Fig.~1.
\end{figure}

\clearpage
\begin{figure}
\plotone{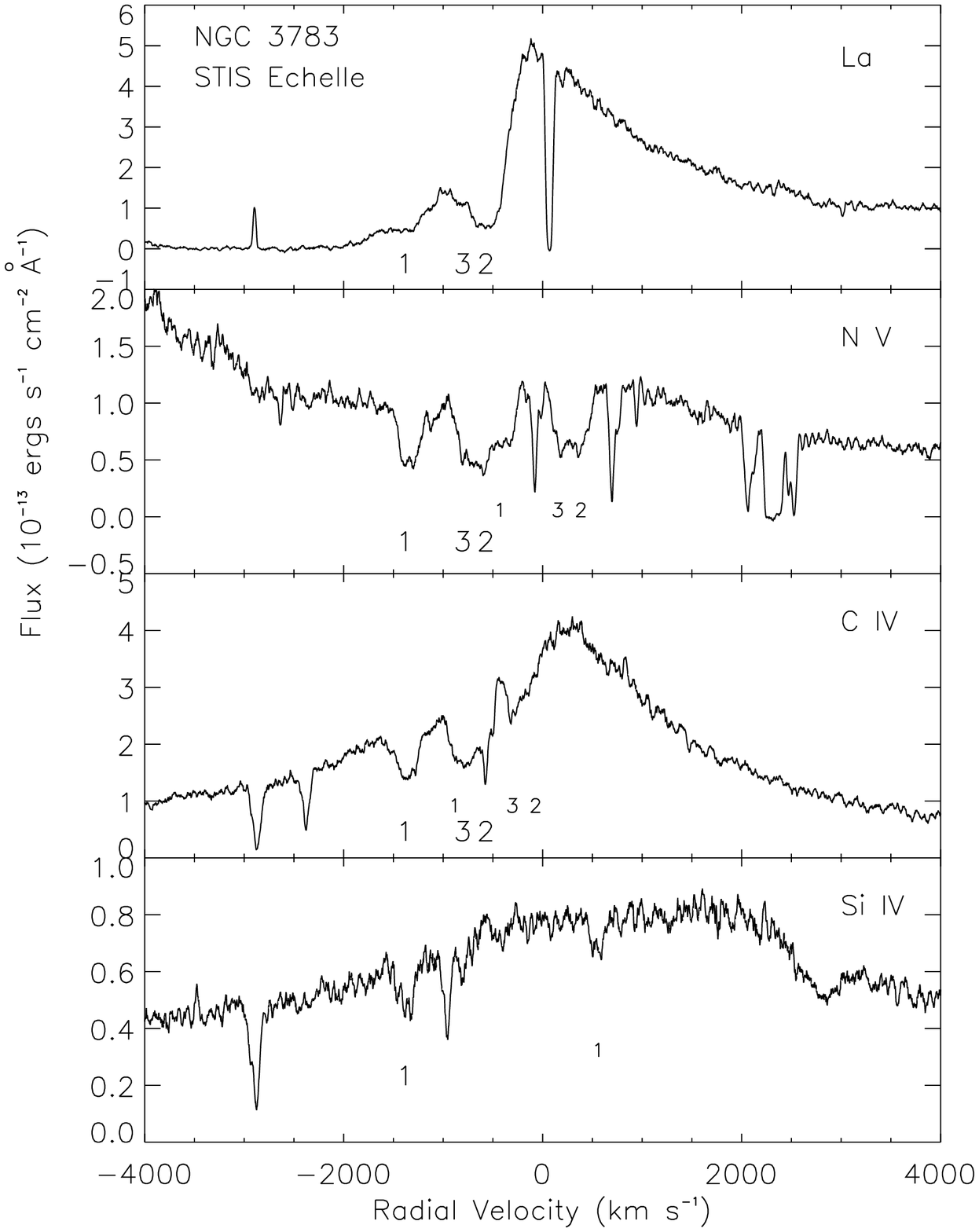}
\\Fig.~2.
\end{figure}

\clearpage
\begin{figure}
\plotone{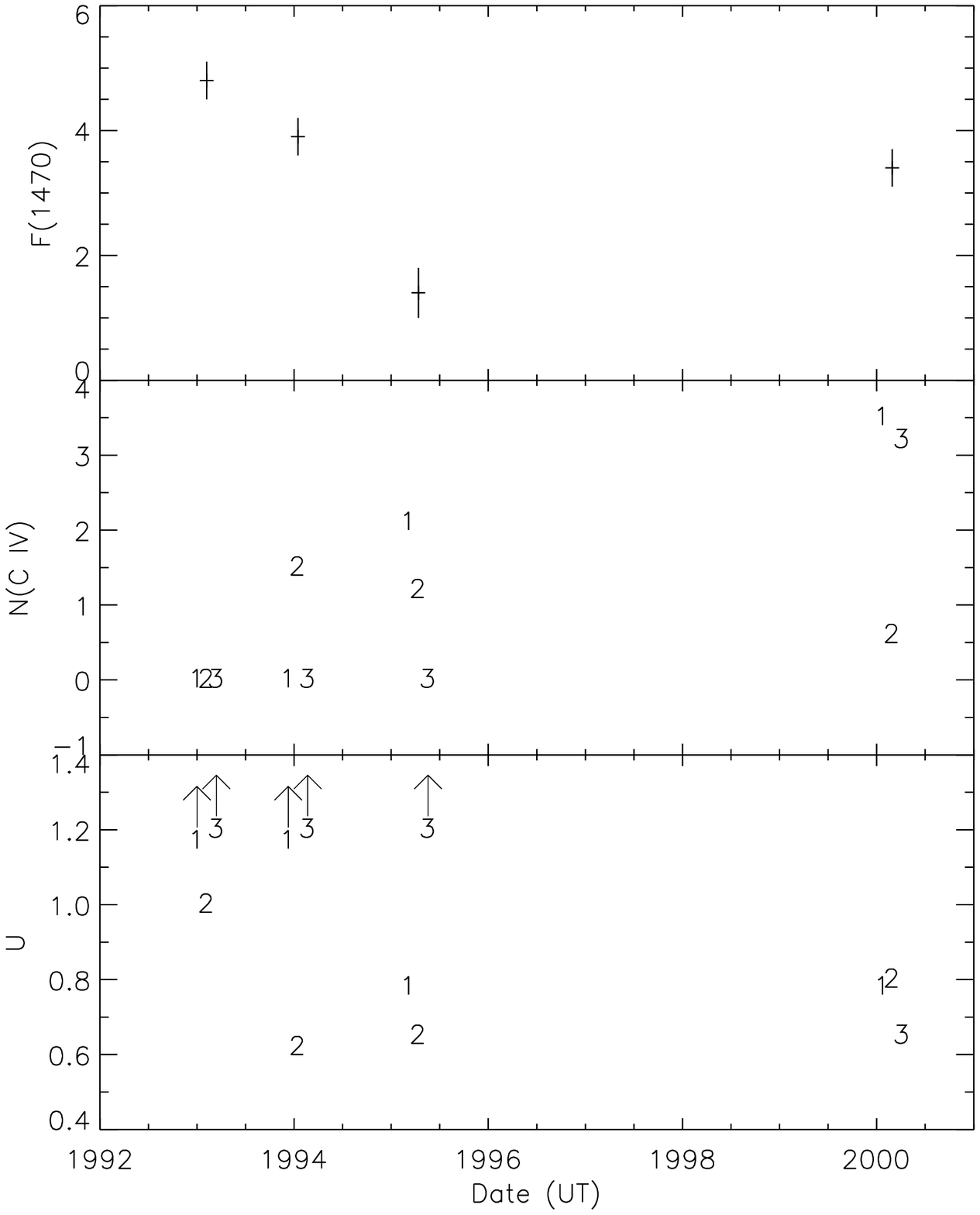}
\\Fig.~3.
\end{figure}

\end{document}